\documentclass{article}
\usepackage{mathptmx}
\usepackage{amsmath}
\usepackage{amssymb}
\usepackage{hyperref}
\usepackage{graphicx}
\usepackage{ulem}
\usepackage{setspace}
\hypersetup{pdfborder=false}
\title{Actions for an Hierarchy of Attractive Nonlinear Oscillators Including The Quartic Oscillator in 1+1 Dimensions}
\author{Robert L. Anderson{\footnote{\texttt{\href{mailto:andersonr@hal.physast.uga.edu}{andersonr@hal.physast.uga.edu}}}}\\
\vspace{.025in}\\
Department of Physics and Astronomy\\
\vspace{.025in}\\
University of Georgia\\
\vspace{.025in}\\
Athens, Georgia 30602}
\date{\today}
\begin{document}                  

\maketitle

\begin{abstract}
{\noindent}In this paper, we present an explicit form in terms of end-point data for the classical action $S_{2n}$ evaluated on extremals satisfying the Hamilton-Jacobi equation for each member of a hierarchy of classical non-relativistic oscillators characterized by even power potentials (i.e., attractive potentials $V_{2n}(y_{2n})={\frac{1}{2n}}k_{2n}y_{2n}^{2n}(t)|_{n{\geq}1}$).   The nonlinear quartic oscillator corresponds to $n=2$ while the harmonic oscillator corresponds to $n=1$. 
\end{abstract}

\part{INTRODUCTION}

The linearization map in [1] gives the solution for all members of the hierarchy $V_{2n}(y_{2n})={\frac{1}{2n}}k_{2n}y_{2n}^{2n}(t)|_{n{\geq}1}$ in terms of the linear (harmonic) oscillator ($n=1$).  It consists of an explicit nonlinear deformation of coordinates and a nonlinear deformation of time coordinates involving a quadrature:

\begin{align*}
x({\hat{t}})&=(k_{2n}/nk_{2})^{1/2}y_{2n}(t)(y_{2n}^{2}(t))^{(n-1)/2},\\
y_{2n}(t)&=(nk_{2}/k_{2n})^{1/2n}x({\hat{t}})(x^{2}({\hat{t}}))^{(1/2)(1-n)/n}, \tag{1.1}    
\end{align*}

\begin{align*}
{\frac{dt}{d{\hat{t}}}}=n^{-(2n-1)/2n}(k_{2}/k_{2n})^{1/2n}(x^{2}({\hat{t}}))^{-(n-1)/2n},\\
{\frac{d{\hat{t}}}{dt}}={\sqrt{n}}(k_{2n}/k_{2})^{1/2}(y_{2n}^{2}(t))^{(n-1)/2},\tag{1.2}    
\end{align*}

So, it important to keep in mind that all the quantities in this paper are known in principle as a consequence of (1.1) and (1.2)!\\
     	
However, it is a non-canonical map.  Therefore, to find a form for the action evaluated on an extremal, we can only use the linearization map as a guide, albeit an extremely useful one.\\

Critical to the form stated below is the fact that $y_{{2n}_{\max}}$ and $t_{\max}$ are constants of the motion and they uniquely characterize every extremal of the periodic systems studied in this paper.  This allows us to find a form in terms of end-point data for the classical action $S_{2n}$ evaluated on extremals for each member of a hierarchy of classical non-relativistic oscillators characterized by even power potentials (i.e.,attractive potentials $V_{2n}(y_{2n})={\frac{1}{2n}}k_{2n}y_{2n}^{2n}(t)|_{n{\geq}1}$).  (The quartic oscillator corresponds to $n=2$ while the harmonic oscillator corresponds to $n=1$. The form is new for the harmonic linear oscillator and using the material in Part II one can readily check that it is equal in value to the one given in [2]-[3].) (See Appendix B.) \\

In particular, we arrive at a form for

\begin{align*}
S_{(2n)}(t_a,y_{{2n}_{a}};t_b,y_{{2n}_{b}})={\overset{t_b}{\underset{t_{\max}}{\int}}}L_{2n}(y_{2n}(t),{\frac{d}{dt}}y_{2n}(t))dt|_{extremal}{\hspace{6pt}},\\
+{\overset{t_{\max}}{\underset{t_{a}}{\int}}}L_{2n}(y_{2n}(t),{\frac{d}{dt}}y_{2n}(t))dt|_{extremal}{\hspace{6pt}},\tag{1.3}    
\end{align*}

where $L_{2n}$ equals the Lagrangian for the $2n$ oscillator, which satisfies

\begin{align*}
{\frac{{\partial}S_{2n}}{{\partial}y_{{2n}_b}}}&=p_{{2n}_b}{\hspace{6pt}},\\
{\frac{{\partial}S_{2n}}{{\partial}t_b}}&=-E=-H_{2n}({\frac{{\partial}S_{2n}}{{\partial}y_{{2n}_b}}},V_{2n}(y_{{2n}_b})){\hspace{6pt}} ,\tag{1.4}    
\end{align*}

and the time-reversed motion

\begin{align*}
{\frac{{\partial}S_{2n}}{{\partial}y_{{2n}_a}}}&=-p_{{2n}_a}{\hspace{6pt}},\\
{\frac{{\partial}S_{2n}}{{\partial}t_a}}&=+E=H_{2n}({\frac{{\partial}S_{2n}}{{\partial}y_{{2n}_a}}},V_{2n}(y_{{2n}_a})){\hspace{6pt}} ,\tag{1.5}    
\end{align*}

where ${H_{2n}}$ = Hamiltonian for the $2n$ oscillator.
                                                    
We shall use the notation $y=y_{2n}$ where it does not cause confusion.

\begin{align*}
&S_{2n}(t_a,y_a;t_b,y_b)  ={\frac{m{\omega}}{(n+1)}}({\frac{k_{2n}}{nk_2}})^{1/2}{\hspace{411pt}}\\
&{\lbrace}[(y^2_b)^{(n+1)/2}{\cos}{\omega}{\int}^{t_b}_{t{_{\max}}}{\gamma}(t)dt-(n+1)y_b{\hspace{2pt}}y_{\max}(y_{{\max}^2})^{(n-1)/2}+n(y^2_{{\max}})^{(n+1)/2}  ({\cos}^2{\omega}{\int}^{t_b}_{t{_{\max}}}{\gamma}(t)dt)^{(n+1)/2n}/{\cos}({\omega}{\int}^{t_b}_{t{_{\max}}}{\gamma}(t)dt)]\\
&/{\sin}({\omega}{\int}^{t_b}_{t{_{\max}}}{\gamma}(t)dt)\\
&+{\hspace{540pt}}\\
&[(y^2_a)^{(n+1)/2}{\cos}{\omega}{\int}^{t_{\max}}_{t{_{a}}}{\gamma}(t)dt-(n+1)y_ay_{\max}(y_{{\max}^2})^{(n-1)/2}+n(y^2_{{\max}})^{(n+1)/2}({\cos}^2{\omega}{\int}^{t_{\max}}_{t{_{a}}}{\gamma}(t)dt)^{(n+1)/2n}/{\cos}({\omega}{\int}^{t_{\max}}_{t{_{a}}}{\gamma}(t)dt)]\\
&/{\sin}({\omega}{\int}^{t_{\max}}_{t{_a}}{\gamma}(t)dt){\rbrace}\\
&+{\hspace{540pt}}\\
&{\frac{(n-1)}{(n+1)}}{\frac{k_{2n}}{2n}}y^{2n}(t_b-t_a).\tag{1.6}    
\end{align*}

where ${\gamma}(t)$ = $({\frac{nk_{2n}}{k_2}})^{1/2}$ $(y^2(t))^{n-1/2}$.\\

In Part II,  we discuss the description of the hierarchy extremals using the known results for the harmonic (linear) oscillator ($n=$1) that are implied by the linearization map presented in [1]. We start with the quartic oscillator and then present the general case.\\

In Part III, we set $n=2$ in (1.6) and show that the classical action evaluated on an extremal $S_{qo}$ satisfies the Hamilton-Jacobi equation equations (1.4)-(1.5)$|_{n=2}$ for the quartic oscillator.\\

In Part IV, we take $n$ arbitrary in (1.6) and show that the classical action evaluated on an extremal $S_{2n}$ satisfies the Hamilton-Jacobi equations (1.4)-(1.5) for each member of the hierarchy. This reproduces the quartic oscillator result in Part  III and as well as yielding a new form for the harmonic oscillator.\\

\part{Extremals}

The set of extremals for the harmonic (linear) oscillator is described by the endpoint solution of Newton's equation of motion

\begin{align*}
x({\hat{t}})=(x_b{\hspace{2pt}}{\sin}{\omega}({\hat{t}}-{\hat{t}}_a)+x_a{\sin}{\omega}({\hat{t}}_b-{\hat{t}})/{\sin}{\omega}({\hat{t}}_b-{\hat{t}}_a), \tag{2.1}    
\end{align*}

where the spring constant $k_2$ = $m{\omega}^2$, $m=mass$, ${\hat{t}}$ denotes the harmonic oscillator time and $x$ denotes the space coordinate of the linear oscillator (ref [3] and eq. (1.1) in [1]).  It is important to note that each extremal is uniquely characterized by $x_{\max}$ and a ${\hat{t}}_{{\max}}$.  Specifically, we take $0{\hspace{2pt}}{\leq}{\hspace{3pt}}{\hat{t}}_{{\max}}<{\frac{2{\pi}}{{\omega}}}$ {\hspace{2pt}} and of course $x_{\max}$ is fixed by the energy. In practice the evaluation of quantities here we can take ${\hat{t}}_{{\max}}$ up to a multiple of the period.

Now the extremals are also described by the equation

\begin{align*}
x({\hat{t}})=x_{\max}{\hspace{2pt}}{\cos}{\omega}({\hat{t}}-{\hat{t}}_{\max}) \tag{2.2}    
\end{align*}

[(see Appendix A for a demonstation)].
 
The set of extremals for a quartic oscillator with mass $m$ is described by the equivalent endpoint solution of Newton's equation of motion

\begin{align*}
&(y^2(t))^{1/2}{\hspace{2pt}}y(t)=\\
&{\Biggl{[}}{\frac{y_b(y^2_b)^{(1/2)}{\sin}{\omega}{\overset{t}{\underset{t_{a}}{\int}}}({\frac{2k_{4}}{k_2}})^{1/2}(y^2(t'))^{1/2}dt'{\hspace{4pt}}+{\hspace{4pt}}y_a(y{_a^2})^{1/2}{\sin}{\omega}{\overset{t_{b}}{\underset{t}{\int}}}({\frac{2k_{4}}{k_2}})^{1/2}(y^2(t'))^{1/2}dt'}{{\sin}{\omega}{\overset{b}{\underset{a}{\int}}}({\frac{2k_{4}}{k_2}})^{1/2}(y^2(t'))^{1/2}dt'}}{\Biggr{]}} \tag{2.3}    
\end{align*}

and the equivalent integral equation

\begin{align*}
(y(t)(y^2(t))^{1/2})-(y_{\max}(y^2_{\max})^{1/2}){\cos}({\omega}{\int}^{t}_{t{_{max}}}{\gamma}(t)dt')=0\tag{2.4}    
\end{align*}

where $\gamma$ = (2$k_4/k_2)^{1/2}(y^2)^{1/2},{\hspace{4pt}}y_{max}$ = $(4E/k_4)^{1/4}$, $k_4$ = denotes the quartic spring constant, $t$ denotes quartic oscillator time and $y=y_{qo}=y_4$ denotes the space coordinate of the quartic oscillator.\\

(We remind the reader that because of the linearization map given in [1], the integral equation (2.4) is solved.  However, as an aside, we would like to point out that ${\int}^{t}_{t{_{max}}}{\gamma}(t')dt'$ can also be determined from (2.4) since we know the pairs $(t,y(t))$ and $(t_{\max}, y_{\max})$.)\\

The argument in Appendix A generalizes to this case for the equivalence of (2.3) and (2.4).  Further, that the integral equation (2.4) satisfies Newton’s equation of motion ${\frac{d^2}{dt^2}}my(t)$ = $-k_4y^3(t)$ follows by direct differentiation twice.\\

The pairs (2.1)-(2.2) and (2.3)-(2.4) are connected by the linearization map (1.1)-1.2) given in [1] between the linear oscillator and the quartic oscillator. \\
                                       
Using the preceding arguments, (2.3)-(2.4) generalize to

\begin{align*}
&(y^2_{2n}(t))^{(n-1)/2}{\hspace{2pt}}y_{2n}(t)=\\
&{\Biggl{[}}{\frac{y_{2n_b}(y_{2n_b}^2)^{(n-1)/2}{\sin}{\omega}{\overset{t}{\underset{t_{a}}{\int}}}{\gamma}(t')dt'{\hspace{4pt}}+{\hspace{4pt}}y_{2n_a}(y_{2n_a}^2)^{(n-1)/2}{\sin}{\omega}{\overset{t_b}{\underset{t}{\int}}}{\gamma}(t')dt'}{{\sin}{\omega}{\overset{t_b}{\underset{t_a}{\int}}}{\gamma}(t')dt'}}{\Biggr{]}} \tag{2.5}    
\end{align*}

which is equivalent to the integral equation 

\begin{align*}
(y_{2n}(t)(y^{2}_{2n}(t))^{(n-1)/2})-(y_{{2n}_{\max}}(y^2_{2n_{\max}})^{(n-1)/2}){\cos}({\omega}{\int}^{t}_{t{_{\max}}}{\gamma}(t')dt')=0,\tag{2.6}    
\end{align*}

where ${\gamma}(t')$ = $({\frac{nk_{2n}}{k_2}})^{1/2}(y_{2n}(t'))^{(n-1)/2}$.\\

The above imply the following momenta since all systems have the same mass $m$:

\begin{align*}
p_{{2n}_b}={\frac{m{\omega}}{ }}({\frac{k_{2n}}{nk_2}})^{1/2}{\bigl{[}}(y^2_{{2n}_b})^{(n-1)/2}y_{{2n}_b}{\cos}({\omega}{\int}^{t_b}_{t{_{\max}}}{\gamma}(t)dt)-(y^2_{{2n}_{\max}})^{(n-1)/2}y_{{2n}_{\max}}{\bigr{]}}/{\sin}({\omega}{\int}^{t_b}_{t{_{\max}}}{\gamma}(t)dt),\tag{2.7a} 
\end{align*}

\begin{align*}
p_{{2n}_a}={\frac{m{\omega}}{ }}({\frac{k_{2n}}{nk_2}})^{1/2}{\bigl{[}}(y^2_{{2n}_a})^{(n-1)/2}y_{{2n}_a}{\cos}({\omega}{\int}^{t_{\max}}_{t{_a}}{\gamma}(t)dt)-(y^2_{{2n}_{\max}})^{(n-1)/2}y_{{2n}_{\max}}{\bigr{]}}/{\sin}({\omega}{\int}^{t_{\max}}_{t{_a}}{\gamma}(t)dt),\tag{2.7b} 
\end{align*}

These are the equations we have to reproduce with our ${S_{2n}}$.\\
\\
\\

Note (2.6) implies
\begin{align*}
(y^2_{2n}(t_b))^{(n-1)/2})=(y^{2}_{{2n}_{\max}})^{(n-1)/2})({\cos}^2({\omega}{\int}^{t_b}_{t{_{\max}}}{\gamma}(t)dt))^{(n-1)/2n},\tag{2.8a}
\end{align*}
and
\begin{align*}
(y^2_{2n}(t_a))^{(n-1)/2})=(y^{2}_{{2n}_{\max}})^{(n-1)/2})({\cos}^2({\omega}{\int}^{t_{\max}}_{t{_a}}{\gamma}(t)dt))^{(n-1)/2n}.\tag{2.8b}    
\end{align*}

These latter relations are needed to evaluate ${\gamma}$ when we differentiate w.r.t time for $n=2$ in Part III and arbitrary $n$ in Part IV.
          
\part{Nonlinear action for the quartic oscillator}
                                        
As mentioned above, because the linearization map given in [1] is a non-canonical one, we do not have a derivation of the actions for the nonlinear quartic oscillator in terms of $y_b,y_a,t_b$ and $t_a$, rather we have constructed it, namely (1.6) with $n=2$, using the linearization map as a guide. \\

\begin{align*}
&S_{qo}(t_a,y_a;t_b,y_b)=S_4(t_a,y_a;t_b,y_b,)\\
&={\frac{m{\omega}}{3}}({\frac{k_4}{2k_2}})^{1/2}\\
&{\lbrace}{\Biggl{[}}(y^2_b)^{3/2}{\cos}({\omega}{\int}^{t_b}_{t_{\max}}{\gamma}(t)dt)-3y_b{\hspace{4pt}}y_{\max}(y^2_{\max})^{1/2}{\hspace{4pt}}+{\hspace{4pt}}2(y^2_{\max})^{3/2}({\cos}^2({\omega}{\int}^{t_b}_{t_{\max}}{\gamma}(t)dt))^{3/4}{\hspace{4pt}}/{\cos}({\omega}{\int}^{t_b}_{t_{\max}}{\gamma}(t)dt)){\Biggr{]}}\\
&/{\sin}({\omega}{\int}^{t_b}_{t_{\max}}{\gamma}(t)dt){\hspace{360pt}}\\
&+\\
&{\Biggl{[}}(y^2_a)^{3/2}{\cos}({\omega}{\int}^{t_{\max}}_{t_a}{\gamma}(t)dt)-3y_a{\hspace{4pt}}y_{\max}(y^2_{\max})^{1/2}{\hspace{4pt}}+{\hspace{4pt}}2(y^2_{\max})^{3/2}({\cos}^2({\omega}{\int}^{t_{\max}}_{t_a}{\gamma}(t)dt))^{3/4}{\hspace{4pt}}/{\cos}({\omega}{\int}^{t_{\max}}_{t_a}{\gamma}(t)dt)){\Biggr{]}}\\
&/{\sin}({\omega}{\int}^{t_{\max}}_{t_a}{\gamma}(t)dt){\rbrace}\\
&+{\frac{1}{12}}{\hspace{2pt}}k_4{\hspace{2pt}}y^4_{\max}(t_b-t_a)\tag{3.1} 
\end{align*}

This implies

\begin{align*}
&{\frac{\partial}{{\partial}y_b}}S_{qo}(t_a,y_a;t_b,y_b)\\
&=m{\omega}({\frac{k_4}{2k_2}})^{1/2}\\
&{\Biggl{[}}(y^2_b)^{1/2}{\hspace{4pt}}y_b{\hspace{4pt}}{\cos}({\omega}{\int}^{t_b}_{t_{\max}}{\gamma}(t)dt)-(y^2_{\max})^{1/2}y_{max}{\Biggr{]}}/{\sin}({\omega}{\int}^{t_b}_{t_{\max}}{\gamma}(t)dt)=p_b\tag{3.2} \\
\end{align*}
\\
\\
and
\begin{align*}
&{\frac{\partial}{{\partial}t_b}}S_{qo}(t_a,y_a;t_b,y_b)\\
&={\frac{m{\omega}^2}{3}}({\frac{k_4}{2k_2}})^{1/2}{\gamma}_b{\lbrace}\\
&{\Biggl{[}}(y^2_b)^{3/2}{\cos}({\omega}{\int}^{t_b}_{t_{\max}}{\gamma}dt)-3y_b{\hspace{4pt}}y_{\max}(y^2_{\max})^{1/2}{\hspace{4pt}}+{\hspace{4pt}}2(y^2_{\max})^{3/2}({\cos}^2({\omega}{\int}^{t_b}_{t_{\max}}{\gamma}(t)dt)^{3/4}{\hspace{4pt}}/{\cos}({\omega}{\int}^{t_b}_{t_{\max}}{\gamma}(t)dt)){\Biggr{]}}\\
&(-{\cos}({\omega}{\int}^{t_b}_{t_{\max}}{\gamma}(t)dt)/{\sin}^2({\omega}{\int}^{t_b}_{t_{\max}}{\gamma}(t)dt))\\
&+[(y^2_b)^{3/2}(-{\sin}({\omega}{\int}^{t_b}_{t_{\max}}{\gamma}(t)dt)\\
&+2(y^2_{\max})^{3/2}(3/4){\langle}-2{\cos}({\omega}{\int}^{t_b}_{t_{\max}}{\gamma}(t)dt){\sin}({\omega}{\int}^{t_b}_{t_{\max}}{\gamma}(t)dt){\rangle}/(({\cos}^2({\omega}{\int}^{t_b}_{t_{\max}}{\gamma}(t)dt))^{1/4}{\cos}({\omega}{\int}^{t_b}_{t_{\max}}{\gamma}(t)dt))\\
&+2(y^2_{\max})^{3/2}({\cos}^2({\omega}{\int}^{t_b}_{t_{\max}}{\gamma}(t)dt))^{3/4}{\sin}({\omega}{\int}^{t_b}_{t_{\max}}{\gamma}(t)dt)/{\cos}^2({\omega}{\int}^{t_b}_{t_{\max}}{\gamma}(t)dt)]\\
&/{\sin}({\omega}{\int}^{t_b}_{t_{\max}}{\gamma}(t)dt){\rbrace}\\
&+{\frac{1}{12}}{\hspace{2pt}}k_4{\hspace{2pt}}y^4_{\max}\\
&=-{\frac{k_4}{3}}y^4_{\max}+{\frac{1}{12}}{\hspace{2pt}}k_4{\hspace{2pt}}y^4_{\max}=-{\frac{k_{4}{\hspace{2pt}}y^4_{\max}}{4}}\tag{3.3} 
\end{align*}

(We have used (2.4) $|_{t=t_b}$ and (2.8a) $|_{n=2}$  after differentiating wrt $t_b$ to obtain (3.3).)  The first four terms sum to zero.                                                                                      

\begin{align*}
{\frac{{\partial}S_{qo}}{{\partial}y_{_b}}}&=p_{{qo}_b}{\hspace{6pt}},\\
{\frac{{\partial}S_{qo}}{{\partial}t_b}}&=-E=H_{qo}({\frac{{\partial}S_{qo}}{{\partial}y_{_b}}},V_{qo}(y_b)){\hspace{6pt}}.\tag{3.4}    
\end{align*}
\\
\\
\\
This agrees with (2.7a) and (1.4) for $n=2$.\\

The $a$-differentiations parallel the $b$-differentiations and yield
\begin{align*}
{\frac{{\partial}S_{qo}}{{\partial}y_{_a}}}&=-p_{{qo}_a}{\hspace{6pt}},\\
{\frac{{\partial}S_{qo}}{{\partial}t_a}}&=+E=H_{qo}({\frac{{\partial}S_{qo}}{{\partial}y_{_a}}},V_{qo}(y_a)){\hspace{6pt}}.\tag{3.5}    
\end{align*}

This agrees with (2.7b) and (1.5) for $n=2$.\\

It is important to note that the value of $S_{qo}$  is not changed if the substitution of (2.4)$|_{t=t_b}$ is made in (3.1).  However, this substitution cannot be made before all differentiations are made because the choice of the form in terms of space-time endpoints for $S_{qo}$  is critical.

\part{Actions for the nonlinear hierarchy $V_2(y_{2n})={\frac{1}{2_n}}k_{2n}y_{2n}^{2n}(t){\hspace{4pt}}|_{n{\geq}1^.}$}

Here we generalize the approach from Part III.

The action on an extremal $S_{2n}(t_a,y_a;t_b,y_b)$ is given (1.4) (Here, we shall use the notation $y=y_{2n}$.).  We now proceed to verify that (1.6) satisfies the Hamilton-Jacobi equations (1.4)-(1.5) by verifying that the following equations are satisfied (We shall use the notation $y=y_{2n}$.): 

\begin{align*}
&{\frac{\partial}{{\partial}y_b}}S_{2n}(t_a,y_a;t_b,y_b)\\
&={\hspace{4pt}}m{\omega}({\frac{k_{2n}}{nk_2}})^{1/2}\\
&{\Biggl{[}}(y^2_b)^{(n-1)/2}{\hspace{4pt}}y_b{\hspace{4pt}}{\cos}({\omega}{\int}^{t_b}_{t_{\max}}{\gamma}(t)dt)-y_{\max}(y^2_{\max})^{(n-1)/2}{\Biggr{]}}\\
&/{\sin}({\omega}{\int}^{t_b}_{t_{\max}}{\gamma}(t)dt)=p_{{2n}_b},\tag{4.1}
\end{align*}   

and

\begin{align*}
&{\frac{\partial}{{\partial}t_b}}S_{2n}(t_a,y_a;t_b,y_b)={\frac{m{\omega}^2}{(n+1)}}({\frac{k_{2n}}{nk_2}})^{1/2}{\gamma}_b\\
&{\lbrace}{\Biggl{[}}(y^2_b)^{(n+1)/2}{\cos}({\omega}{\int}^{t_b}_{t_{\max}}{\gamma}(t)dt-(n+1)y_b{\hspace{4pt}}y_{\max}(y^2_{\max})^{(n-1)/2}{\hspace{4pt}}+{\hspace{4pt}}n(y^2_{\max})^{(n+1)/2}({\cos}^2({\omega}{\int}^{t_b}_{t_{\max}}{\gamma}(t)dt))^{(n+1)/2n}/{\cos}({\omega}{\int}^{t_b}_{t_{\max}}{\gamma}(t)dt){\Biggr{]}}\\
&(-{\cos}({\omega}{\int}^{t_b}_{t_{\max}}{\gamma}(t)dt)/{\sin}^2({\omega}{\int}^{t_b}_{t_{\max}}{\gamma}(t)dt)\\
&+{\Biggl{[}}(y^2_a)^{(n+1)/2}-{\sin}({\omega}{\int}^{t_b}_{t_{\max}}{\gamma}(t)dt))+n(y^2_{\max})^{(n+1)/2n}((n+1)/2){\cos}^2({\omega}{\int}^{t_b}_{t_{\max}}{\gamma}(t)dt)^{((n+1)/2n-1}{\langle}-2{\cos}({\omega}{\int}^{t_b}_{t_{\max}}{\gamma}(t)dt)\\
&{\sin}({\omega}{\int}^{t_b}_{t_{\max}}{\gamma}(t)dt){\rangle}/{\cos}({\omega}{\int}^{t_b}_{t_{\max}}{\gamma}(t)dt)+n(y^2_{\max})^{(n+1)/2}{\cos}^2{\omega}{\int}^{t_b}_{t_{\max}}{\gamma}(t)dt)^{(n+1)/2n}{\hspace{4pt}}{\sin}({\omega}{\int}^{t_b}_{t_{\max}}{\gamma}(t)dt)/({\cos}^2({\omega}{\int}^{t_b}_{t_{\max}}{\gamma}(t)dt.){\Biggr{]}}\\
&/{\sin}({\omega}{\int}^{t_b}_{t_{\max}}(t)dt){\rbrace}\\
&+{\frac{n-1}{n+1}}{\hspace{2pt}}{\frac{k_{2n}}{2n}}{\hspace{2pt}}y^{2n}_{\max}\\
&=-{\frac{k_{2n}}{n+1}}y^{2n}_{\max}+{\frac{(n-1)}{(n+1)}}{\frac{k_{2n}}{2n}}y^{2n}_{\max},\tag{4.2} 
\end{align*}

where ${\gamma}_b$ = $({\frac{nk_{2n}}{k_2}})^{1/2}(y^{2}_{b})^{(n-1)/2}$.\\

(We have used (2.4) $|_{t=t_b}$ and (2.8a) after differentiating $t_b$ to obtain (4.2).)    The first four terms sum to zero.

Thus,
\begin{align*}
{\frac{{\partial}S_{2n}}{{\partial}y_b}}&=p_{{2n}_b}{\hspace{3pt}},\\
{\frac{{\partial}S_{2n}}{{\partial}t_b}}&=-E,{\hspace{3pt}}\tag{4.3}    
\end{align*} 

which agrees with (2.7a) and (1.4).\\

The $a$-differentiations parallel the $b$-differentiations and yield.

\begin{align*}
{\frac{{\partial}S_{2n}}{{\partial}y_a}}&=p_{{2n}_a}{\hspace{3pt}},\\
{\frac{{\partial}S_{2n}}{{\partial}t_a}}&=E,{\hspace{3pt}}\tag{4.4}    
\end{align*} 

which agrees with (2.7b) and (1.5).\\

\part{Concluding Remarks}

To the best of the author's knowledge of the existing literature on classical mechanics, he can not find any literature on general transformation theory devoted to transforming one system at a given time to another at a distinctly different time, hence the absence of references to the classical literature on this point in this paper. \\

\part{Acknowledgements}

The author wishes to acknowledge those who participated in a seminar organized by David Edwards and Robert Varley in AY 2006-2007 to study Feynman Path Integrals and especially two students Emily Pritchett and Justin Manning.  The seminar provided the original motivation for exploring the extent of  the connection between the linear oscillator and the Feynman’s Path Integral Method of which this is a part. Further, Robert Varley has continued to this date to provide insight and encouragement in this effort.\\

 The author wishes to thank Professor Howard Lee for his insightful discussions and his constant encouragement. \\   

\section*{References}

[1] Robert L. Anderson, “An Invertible linearization map for the quartic oscillator”, JMP {\textbf{51}} , 122904 (2010).

[2] Feynman’s Thesis-A New Approach To Quantum Theory, Laurie M. Brown-editor. Singapore: World Scientific, 2005.  See equation (27), p 18 and set . We are grateful to Justin Manning for bringing this result to our attention.  

[3] R. P. Feynman and A. R. Hibbs, Quantum Mechanics and Path Integrals, McGraw-Hill, 1965.

\newpage

\section*{Appendix A}

As stated in Part II, the set of extremals for the harmonic (linear) oscillator is given by\\
 $x({\hat{t}})=(x_b{\hspace{2pt}}{\sin}{\omega}({\hat{t}}-{\hat{t}}_a)+x_a{\sin}{\omega}({\hat{t}}_b-{\hat{t}})/{\sin}{\omega}({\hat{t}}_b-{\hat{t}}_a)$, ${\hspace{45pt}}$(2.1)   

where the spring constant $k_2$ = $m{\omega}^2$, $m$ = $mass$, ${\hat{t}}$ denotes the harmonic oscillator time and $x$ denotes the space coordinate of the linear oscillator.    

We now demonstrate that equation (A.1) is equivalent to the relationship\\
$x({\hat{t}})=x_{\max}{\hspace{2pt}}{\cos}{\omega}({\hat{t}}-{\hat{t}}_{\max})$.${\hspace{150pt}}$ (2.2)   

Rewriting (2.1) as

\begin{align*}
x({\hat{t}})=(x_b{\hspace{2pt}}{\sin}{\omega}({\hat{t}}-{\hat{t}}_{\max}+{\hat{t}}_{\max}-{\hat{t}}_a)+x_a{\sin}{\omega}({\hat{t}}_b-{\hat{t}}_{\max}+{\hat{t}}_{\max}-{\hat{t}}))/{\sin}{\omega}({\hat{t}}_b-{\hat{t}}_{\max}+{\hat{t}}_{\max}-{\hat{t}}_a), \tag{A.1}    
\end{align*}

cross-multiplying and expanding, we have

\begin{align*}
x({\hat{t}}){\hspace{2pt}}{\sin}{\omega}({\hat{t}}_b-{\hat{t}}_{\max}+{\hat{t}}_{\max}-{\hat{t}}_a)=(x_b{\sin}{\omega}({\hat{t}}-{\hat{t}}_{\max}+{\hat{t}}_{\max}-{\hat{t}}_a)+x_a{\sin}{\omega}({\hat{t}}_b-{\hat{t}}_{\max}+{\hat{t}}_{\max}-{\hat{t}}))    
\end{align*}
${\hspace{50pt}}$or
\begin{align*}
x({\hat{t}}){\lbrace}{\sin}{\omega}({\hat{t}}_b-{\hat{t}}_{\max}){\cos}{\omega}({\hat{t}}_{\max}-{\hat{t}}_a)+{\cos}{\omega}({\hat{t}}_b-{\hat{t}}_{\max}){\sin}{\omega}({\hat{t}}_{\max}-{\hat{t}}_a){\rbrace}\\
=x_b{\lbrace}{\sin}{\omega}({\hat{t}}-{\hat{t}}_{\max}){\cos}{\omega}({\hat{t}}_{\max}-{\hat{t}}_a)+{\cos}{\omega}({\hat{t}}-{\hat{t}}_{\max}){\sin}{\omega}({\hat{t}}_{\max}-{\hat{t}}_a){\rbrace}\\
+x_a{\lbrace}{\sin}{\omega}({\hat{t}}_b-{\hat{t}}_{\max}){\cos}{\omega}({\hat{t}}_{\max}-{\hat{t}})+{\cos}{\omega}({\hat{t}}_b-{\hat{t}}_{\max}){\sin}{\omega}({\hat{t}}_{\max}-{\hat{t}}){\rbrace}\\    \tag{A.2}
\end{align*}

\newpage

Substituting (2.2), we obtain the identity

\begin{align*}
x({\hat{t}}){\hspace{2pt}}{\sin}{\omega}({\hat{t}}_b-{\hat{t}}_{\max}){\frac{x_a}{x_{\max}}}+x({\hat{t}}){\frac{x_b}{x_{\max}}}{\sin}{\omega}({\hat{t}}_{\max}-{\hat{t}}_a)\\
=x_b{\sin}{\omega}({\hat{t}}-{\hat{t}}_{\max}){\frac{x_a}{x_{\max}}}+x_b{\frac{x({\hat{t}})}{x_{\max}}}{\sin}{\omega}({\hat{t}}_{\max}-{\hat{t}}_a)\\
+x_a{\sin}{\omega}({\hat{t}}_b-{\hat{t}}_{\max}){\frac{x({\hat{t}})}{x_{\max}}}+x_a{\frac{x_b}{x_{\max}}}{\sin}{\omega}({\hat{t}}_{\max}-{\hat{t}}),    \tag{A.3}
\end{align*}

where the 1st and the 4th terms on the r.h.s. of (A.3) cancel.\\

You can now run the above argument backwards.  Thus we have shown that (2.1) and (2.2) are equivalent.\\

\section*{Appendix B}

It follows from differentiating (2.1) and (2.2) and setting ${\hat{t}}$ = ${\hat{t}}_b$ that

\begin{align*}
-x_{\max}{\hspace{2pt}}{\sin}{\omega}({\hat{t}}_b-{\hat{t}}_{\max})={\frac{x_b{\cos}{\omega}({\hat{t}}_b-{\hat{t}}_a)-x_a}{{\sin}{\omega}({\hat{t}}_b-{\hat{t}}_a)}}.\tag{B.1}
\end{align*}

Similarly, it follows that

\begin{align*}
x_{\max}{\hspace{2pt}}{\sin}{\omega}({\hat{t}}_{\max}-{\hat{t}}_a)={\frac{x_b-x_a{\cos}{\omega}({\hat{t}}_b-{\hat{t}}_a)}{{\sin}{\omega}({\hat{t}}_b-{\hat{t}}_a)}}.\tag{B.2}
\end{align*}

Hence

\begin{align*}
&m{\omega}[{\frac{(x_b^2+x_a^2){\cos}{\omega}({\hat{t}}_b-{\hat{t}}_a)-2x_bx_a}{2{\sin}{\omega}({\hat{t}}_b-{\hat{t}}_a)}}]\\
=&{\frac{m{\omega}}{2}}[-x_{\max}x_b{\sin}{\omega}({\hat{t}}_b-{\hat{t}}_{\max})-x_{\max}x_a{\sin}{\omega}({\hat{t}}_{\max}-{\hat{t}}_a)].\tag{B.3}
\end{align*}

where the l.h.s. is the $S_{ho}$ of [2] - [3].\\

The action $S_{2n}|_{n=1}=S_{ho}$ with $x=y_2$ given by (1.6) is

\begin{align*}
&S_{ho}={\frac{m{\omega}}{2}}{\lbrace}[x^2_b{\cos}{\omega}({\hat{t}}_b-{\hat{t}}_{\max})-2x_bx_{\max}+{\frac{x^2_{\max}{\cos}^2{\omega}({\hat{t}}_b-{\hat{t}}_{\max})}{{\cos}{\omega}({\hat{t}}_b-{\hat{t}}_{\max})}}]/{\sin}{\omega}({\hat{t}}_b-{\hat{t}}_{\max})\\
&+[x^2_a{\cos}{\omega}({\hat{t}}_{\max}-{\hat{t}}_a)-2x_ax_{\max}+{\frac{x^2_{\max}{\cos}^2{\omega}({\hat{t}}_{\max}-{\hat{t}}_a)}{{\cos}{\omega}({\hat{t}}_{\max}-{\hat{t}}_a)}}]/{\sin}{\omega}({\hat{t}}_{\max}-{\hat{t}}_a)\tag{B.4}
\end{align*}

Now (B.4)$|_{(2.2)}$ in value is given by

\begin{align*}
&{\frac{m{\omega}}{2}}{\lbrace}x^2_b[{\cos}{\omega}({\hat{t}}_b-{\hat{t}}_{\max})-{\frac{1}{{\cos}{\omega}({\hat{t}}_b-{\hat{t}}_{\max})}}]/{\sin}{\omega}({\hat{t}}_b-{\hat{t}}_{\max})\\
&+x^2_a[{\cos}{\omega}({\hat{t}}_{\max}-{\hat{t}}_a)-{\frac{1}{{\cos}{\omega}({\hat{t}}_{\max}-{\hat{t}}_a)}}]/{\sin}{\omega}({\hat{t}}_{\max}-{\hat{t}}_a){\rbrace}\\
&={\frac{m{\omega}}{2}}[-x_bx_{\max}{\sin}{\omega}({\hat{t}}_b-{\hat{t}}_{\max})-x_ax_{\max}{\sin}{\omega}({\hat{t}}_{\max}-{\hat{t}}_a)].\tag{B.5}
\end{align*}

The r.h.s. of (B.5) equals the l.h.s. of (B.3).

\end{document}